# Exploring the Attack Surface of WebSocket

Saeid Ghasemshirazi
Department of Industrial Engineering
Iran University of Science and Technology
Kerman, Iran
saeidgs@yahoo.com

Pouya Heydarabadi
Department of Computer Science
Shahid Bahonar University of Kerman
Kerman, Iran
heydarabadip@gmail.com

*Abstract*—Over the years, with the advancement of technology, Web technology has many improvements. In the early days, the web was one-way communication, and only the customer was able to see the content of the site and could not enter information. However, day by day, the web made significant progress, and technologies such as HTTP, ajax, WebSocket introduced that make pages dynamic and Give us both sides. In short, it is a new type of communications protocol, which was faster and more efficient than previous communication protocols. After the web socket's unveiling, like any other technology, Its security has been discussed, and technology's security has always been a challenge for us. Therefore, in this article, we examine the structure and security problems that can occur in a web socket to choose an excellent alternative to HTTP and use it.

## I. INTRODUCTION

In the past, web pages were static and only meant to display minimal content to interact with the user. Since then, web applications have evolved so much that applications that were on the desktop can now be created on web applications, and communications today are two-way, which is much older than before [1][10][17].

The user is allowed to cooperate in the process and see the answer to a request [3][8]. The biggest thing holding them back was the traditional HTTP model of client-initiated transactions. For instance, watch movies online, receiving and sending emails, or online converters. Browsers play an essential role, and only with advanced browsers, one can use various data and technologies, including WebSocket. When a new technology appears, security issues show up, including WebSocket technology [1][3][6][8][22].

The socket has a two-way connection in web technology and has excellent security and performance in a web application. In this article, we will describe its vulnerabilities step by step [9][17][18].

## II. BACKGROUND

This chapter will learn more about WebSocket technology, connect its history and frame, and upgrade from HTTP to TCP.

### A. WebSocket History

Hypertext Transfer Protocol (HTTP) is an application-layer protocol for collaborative, distributed, hypermedia information systems that allow users to communicate data on the Internet [1]. It was designed for communication between web browsers and web servers, and web application development requires HTTP [20].

The problem is that instead of using HTTP, we use a WebSocket to contact the TCP connection. In 2008, 10 features were created in html5 to improve TCP connection; the Security of API in the WebSocket changed [10][17][20].

Some goal of the WebSocket is as follows:

- Integrates firewalls and routers
- Mutual connections were allowed
- Integrated with Cookie-based authentication
- Merged with HTTP loading scales
- Compatible with binary data

The final version was released in December 2011. A single TCP connection was combined with WebSocket API to replace HTTP polling [20].

### B. WebSocket Performance

A WebSocket is an independent TCP-based TCP protocol; its only connection to HTTP is updating the connection from HTTP to the WebSocket [20][1][19].

Port 80 is used for unencrypted connections, and port 443 is used for secure and encrypted connections [1].

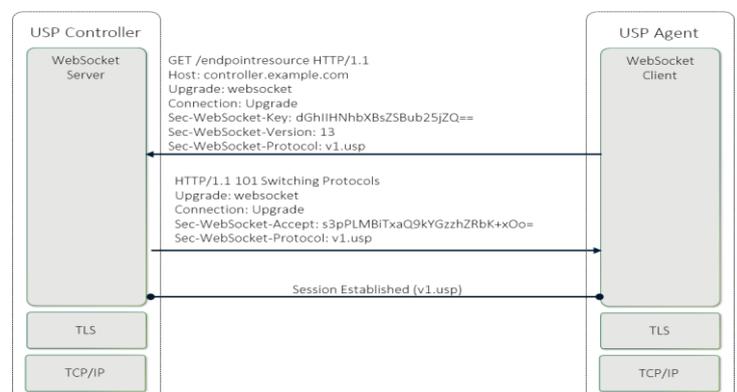

Fig. 1. WebSocket connections workflow

The WebSocket protocol used by web applications starts its connection to the server with HTTP and sends the HTTP request [10][3][20].

This is done in the header upgrade, often called a handshake [3][9]. After agreeing to upgrade the connection with the WebSocket, the client/server can both send data in full-duplex in binary format and depending on the opcode that is used in the web frame socket [21][18][9]. The UTF-8 text is provided. Data converts to binary format or transmits data in binary format without converting and this connection when the handshake is closed.

WebSocket advantages:

- Full dual connection
- Excellent security and low time delay
- Less bandwidth consumption

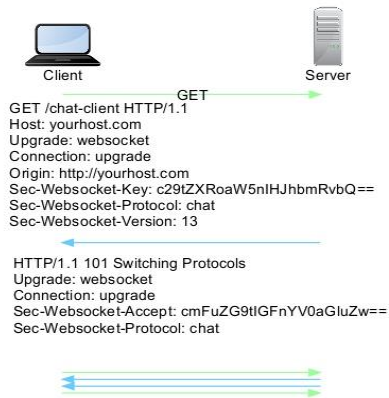

Fig. 2. WebSocket handshake and communication

### C. WebSocket Frame Format

Once the communication channel is established between the WebSocket server and the server, the parties can use the WebSocket.

The WebSocket frame and the connection are entirely two-way. Each participating connection can send the data at the same time [24][23].

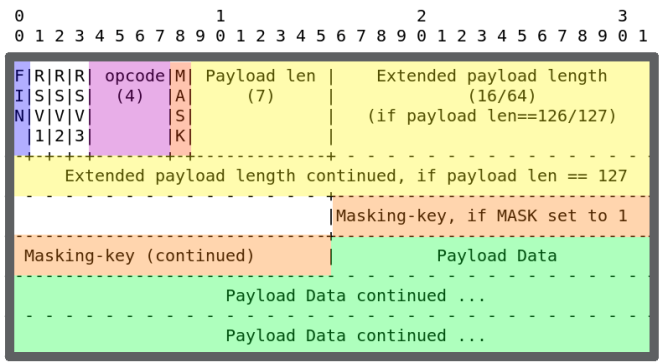

Fig. 3. Web Socket Frame

In many web technologies, sockets are universal. Nevertheless, most security controls are in the hands of the

### III. WEBSOCKET SECURITY

developers. Developers cause vulnerabilities due to lack of time or negligence and lack of sufficient time [4][6][25].

The best solution to eliminate this negligence and reduce vulnerabilities is a security assessment called penetration testing [25]. We look at WebSocket security issues in three sections.

### A. Authentication

Authentication is a way to verify a legitimate user in the service, and the WebSocket does not use a specific authentication method [25][6].

The most common way is the header set-cookie, which can depend on the developer's attitude [25].

Authentication allows credits and licenses to be for resources and what access a person's data can gain.

Authorization is highly dependent on the app-context header. In this case, three states can occur:

- The attacker may be able to achieve a function that does not require a higher-level license from the web service, as confirmed from the beginning
- The hacker can see the content of another user [25]
- The hacker can access anything without permission, and for these reasons, the security of a WebSocket is essential

### B. Same Origin Policy (SOP)

The same source policy is a security method found in most browsers and restrictions [12][11]. Such as documents or scripts (or other data) downloaded from a specific source, restricted and blocked or prevented from execution in communicating [25].

Most web applications use this policy and follow its rules. The rules are such that they work based on a few specific filters:

- Port number
- Different protocol
- Different domains

| url | sop | Explanation |
|---|---|---|
| http://store.company.com/dir2/other.html | yes | the same protocol, host and port. |
| http://store.company.com/dir/inner/another.html | yes | the same protocol, host and port. |
| https://store.company.com/secure.html | no | other protocol. |
| http://store.company.com:81/dir/etc.html | no | other port. |
| http://news.company.com/dir/other.html | no | other host. |

Fig. 4. Types of filtering sop(same origin policy)

The routes inside the authorized domain are not a problem for us. Filtering is more based on the developer's attitude and can determine which method to use and which not to use [5][25]. The way it works is that of HTTP headers [8].

There is a header called Origin, and if its value is set to '*' all domains are allowed [25]. If some domains are specified, only the same domains' data are accepted [15][14].

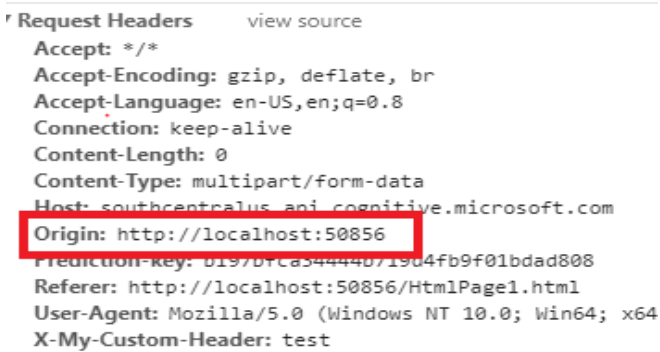

Fig. 5. Filter Origin Header Domain Allow http://localhost:50856 site origin

### C. Cross-Site-Web-Socket-Hijacking(CSWH)

CSWH (Cross-Site-Web-Socket-Hijacking) is a type of CSRF attack. It creates serious security problems and can steal all customer's information. Like other bugs, it is exploitable depending on the hacker's creativity and can do whatever it wants. This attack is more client-side and works with user information [25].

For an attacker to execute his plan, he needs to lead the victim to the malicious page [12][5], so he uses methods such as XSS(Cross-Site-Scripting) or Open-Redirection [5][25][7]. The hacker can creatively send malicious JavaScript code to the victim browser using a hacking tool such as BeEF. If the origin header is uncontrolled, the malicious code is allowed to run. It allows a link to be established with the target site [12][13][25].

Hackers are always looking to make things better, improve, and push until their solution is nearly flawless. Now, the hacker can execute the code he has already written, or the attacker is free to do whatever he wants. This is the main problem, and now he can use the victim's credentials to do whatever he wants [13][16].

Attackers can use this bug in three possible ways:

- Suppose we use a WebSocket to communicate, for instance, in careful operations such as money transactions, realtime applications, and data communication. In that case, the hacker can impersonate the person and start sending sensitive information.

- Suppose we use the WebSocket in places where we can recover our information, such as forgetting the password. In that case, the hacker can make a fake request and start getting sensitive data and, finally, account-takeover.

- The hacker can start sniffing the traffic between the client and the server.

Send the Request and now misuse the input parameters and start sending fake information to the victim browser. It sends the connection request, but before that, it adds the header token. Now hackers can execute JavaScript malware to get access to the client's data [25][5][13].

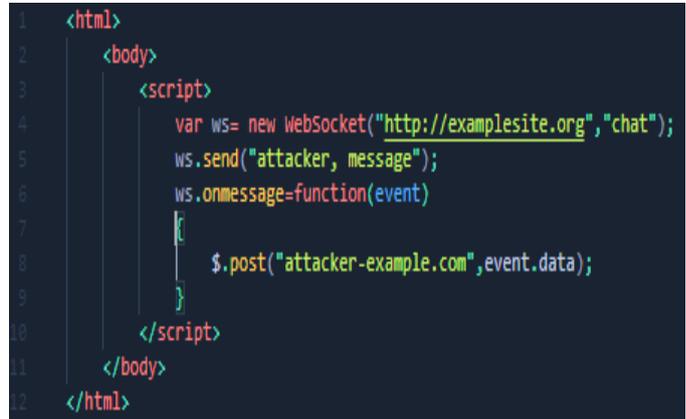

Fig. 6. Cross-Site WebSocket Hijacking (CSWSH) attack scenario

What should we do to fix the vulnerability?

When a handshake occurs, we have to control the inputs and see what data is exchanged.

We especially have to be more careful about the headers' values during the handshake [25].

- We are checking the values of the header origin. As we discussed, the Origin in the previous section prevents untrusted information from other domains. So that data from certain domains are only allowed to enter our domain. We do not have to set the origin header value '*' to send any data to us[]; we must provide a list of valid domains that Origin has access to them.

- In this section should use a method like CSRF-TOKEN [5][25]. We need to generate a token or a random value in the server that the hacker will not connect until it has it. This value will be sent to the server, and it will be allowed to establish if it is correct; if the hacker does not have a valid token, he is not allowed to access, and it is the best solution.

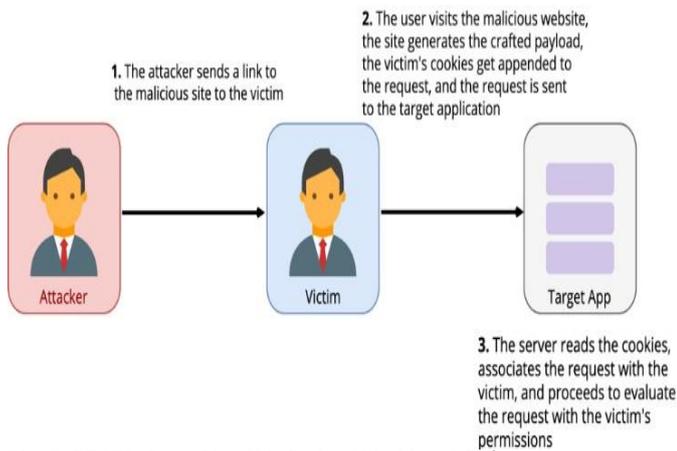

Fig. 7. CSWH (Cross-Site-Web-Socket-Hijacking) POC[1]

### D. Traffic Encryption

Encryption is used to protect the data's confidentiality. When the data is sent, only two parties are aware of it. Without encryption, an attacker can read traffic with MITM (man in the middle). If an attacker can perform a MITM attack, he can read all the information exchanged, which is unacceptable [5][25].

By default, the WebSocket uses port 80, which is non-encrypted. We use port 443 to secure our connections, which is the SSL that saves our information against any MITM attacks [1].

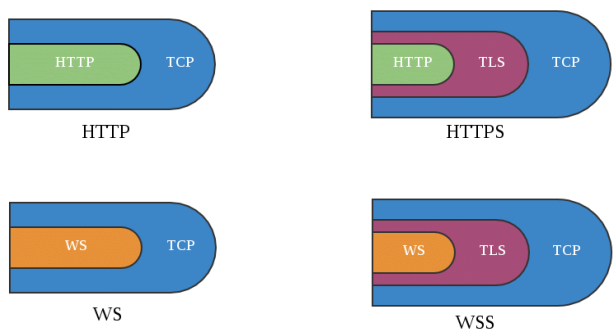

Fig. 8. Secure Connection and Unsecure Connection in WebSocket

### E. Sec-WebSocket-Key

The Sec-WebSocket-Key is a random string with 16 bytes in ASCII range values representing 'nonce' for secure communication.

A client sends 16 bytes of data within a base64 encoded format. The Sec-WebSocket-Accept header is used in the WebSocket opening handshake. This header is sent from server to client to inform that server is willing to initiate a WebSocket connection. The server first receives Sec-WebSocket-Key sent by the client in the first Request and concatenate that with the 'magic string.' Then calculate SHA-1 of the new result and sends base64 of the hashed value.

Sec-WebSocket-Accept:base64(SHA-1(sec-WebSocket-key + "258EAFA5-E914-47DA-95CA-C5AB0DC85B11"))

The 258EAFA5-E914-47DA-95CA-C5AB0DC85B11 is magic string.

In some cases, the programmer does not want to create a better solution for the problem and use magic strings to simplify the testing phase.

### IV. RELATED WORKS

In 2012, an article entitled WebSocket Security Analysis [25] provided information on Web socket vulnerabilities. Increase the web socket's security and productivity in the article and create it as an Explore Surface of Attack of WebSocket.

Our motivation and effort in writing the article were to raise the Socket website's academic level, preventing the website from providing the best possible option for using HTTP.

### V. CONCLUSION

As can see in this article, the WebSocket has much power and makes have the function and speed of information exchange, but we still see HTTP technologies and less use the WebSocket somewhere.

In the near future, most web pages will use web sockets, and this will be achieved when the technology of web sockets reaches a power that can attract everyone's attention and make it look better than HTTP.

Web sockets have not yet been used as a pervasive technology, which is why no other vulnerabilities have been found to find a better platform for its development.

We discussed that if we do not take the existing security problems seriously, We will face serious problems, so we have presented this article to learn more about its security problems and fix them.

---

[1] Proof of concept